\documentclass[aps,prl,showpacs,twocolumn]{revtex4}
\usepackage{graphicx}

\input{epsf}

\begin{document}

\title{ Magnetotransport in the presence of a longitudinal
barrier: multiple quantum interference of edge states}
\author{A. M. Kadigrobov\footnote{E-mail: kadig@tp3.ruhr-uni-bochum.de}$^{1}$, M. V. Fistul$^{1}$, and K. B. Efetov$^{1,2}$
} \affiliation{$^{1}$Theoretische Physik III, Ruhr-Universit\"{a}t
Bochum, D-44801 Bochum,
Germany\\
$^{2}$ L. D. Landau Institute for Theoretical Physics, 117940 Moscow, Russia}
\date{\today}

\begin{abstract}
Transport in a two-dimensional electron gas subject to an
external magnetic field is analyzed in the presence of a \textit{%
longitudinal barrier.} We show that \textit{quantum interference
of the edge states} bound by the longitudinal barrier results in a
drastic change of the electron motion: the degenerate discrete
Landau levels are transformed into an alternating sequence of
energy bands and energy gaps. These features of the electron
spectrum should result in a high sensitivity of thermodynamic and
transport properties of the 2D electron gas to external fields. In
particular, we predict giant oscillations of the ballistic
conductance and discuss nonlinear current-voltage characteristics,
coherent Bloch oscillations and effects of impurities.
\end{abstract}

\pacs{75.47.-m, 05.60.Gg, 75.45.+j, 03.65.Ge
}
\maketitle


Great attention has been attracted during the last decades to
study of transport properties of various mesoscopic systems,
e.g.
ballistic and tunnel junctions, quantum dots,
etc.\cite{MestranspGen1,MestranspGen2}. Such systems display
fascinating quantum-mechanical behavior on a macroscopic
scale, which results in quantization of the conductance \cite%
{MestranspGen1,Beenakker}, Coulomb blockade \cite{MestranspGen2}%
, weak localization \cite{MestranspGen1,MestranspGen2}, mesoscopic
conductance fluctuations \cite{Marcus} and macroscopic quantum tunneling \cite%
{Klarke}, just to name a few. All quantum-mechanical effects are
enhanced in low-dimensional systems, such as a two-dimensional
electron gas, quasi-one-dimensional quantum wires, systems of
coupled small tunnel junctions. Moreover, since the application of
an external magnetic field allows to transform the continuous
spectrum of electrons to discrete Landau levels (in a
two-dimensional electron gas), various quantum-mechanical effects
like Shubnikov-de Haas oscillations \cite{MestranspGen1}, integer
and fractional quantum Hall effects
\cite{MestranspGen1,MestranspGen2}, etc. have been observed in
magnetotransport measurements in these systems.

It is clear that if
a potential barrier is placed across the direction of the electron
motion, the current would flow only due to tunnelling through the
barrier. However, what can happen if the barrier is created along
the direction of the current? To the best of our knowledge this
question has not been addressed yet. Of course, the problem is not
very interesting in the absence of a magnetic field but the
situation drastically changes if a magnetic field is applied
perpendicularly to the plane of the $2D$ electron gas.

In this Letter we show that the quantum-mechanical effects in the
magnetotransport phenomena are enhanced and qualitatively change if such a
``longitudinal'' barrier is present in the system. To be specific, we
consider a two-dimensional electron gas (2DEG)
subject to an external magnetic field $H$. We assume that a
potential barrier of a narrow width separates the systems into two
parts (left and right). What is important, the barrier should be
penetrable, such that electrons can tunnel from one part of the
system to the other. The tunneling through the barrier can
generally be characterized by a reflection amplitude $r$ that can
vary from zero to one.
A schematic setup is shown in Fig. 1a. Here we would like to
emphasize that such a setup is realistic and similar systems have been
produced by using a split-gate technique or cleaved edge fabrication method %
\cite{Technology}.

We start our discussion with a qualitative analysis of the energy spectrum
of the electrons. Effects of the external magnetic field are considered in
the Landau gauge, i.e. the vector-potential $\mathbf{A}%
=\left( -Hy,0,0\right) $, where the axis $y$ is perpendicular to the
barrier. In this gauge the component $p_{x}$ of the momentum conserves even
in the presence of the longitudinal barrier along the $x$%
-axis.

In the limit of a completely transparent barrier, $r=0$, all
states are just Landau levels (the size of the system in the
$y$-direction is assumed to be large), and the electron spectrum
is $\epsilon _{n}(p_{x})~=~\hbar \omega _{c}(n+1/2)$, where
$\omega _{c}~=~eH/mc$ is the cyclotron frequency and $m$ is the
electron  mass. Such a spectrum is shown by dashed lines in Fig.
1b. In this limit  the energy spectrum is independent of $p_{x}$.

In the opposite limit of the zero barrier transparency, $r=1$, the
electron motion near the barrier considerably changes and can be
described in terms of independent \textit{edge states} in the left
and right parts of the system (see Fig. 1a). In this case the
degeneracy of the Landau levels is lifted, and the spectrum of the
edge states near the barrier depends on $p_{x}$. The corresponding
spectrum is represented in Fig. 1b by solid lines. A peculiar
property of such a spectrum is the presence of ``crossing
points'', the number of which grows with an increase of the
quasiclassical parameter $\alpha ~=~\epsilon _{F}/(\hbar \omega
_{c})$, where $\epsilon _{F}$ is the Fermi energy of electrons in
the absence of magnetic field. Indeed, ``the distance'' between
the neighbouring crossing points is $\delta p_{x}~\simeq ~\hbar
/R_{c}$ with $R_c ~=~ c p_F/(e H)$ being the cyclotron radius of
electron orbits. As the momentum $p_{x}$ is restricted by the
Fermi momentum $p_{F}$, the number of the crossing points is
determined by $p_{F}/\delta p_{x}~=~\alpha \gg 1$.

\begin{figure}[tbp]
\includegraphics[width=2.5in]{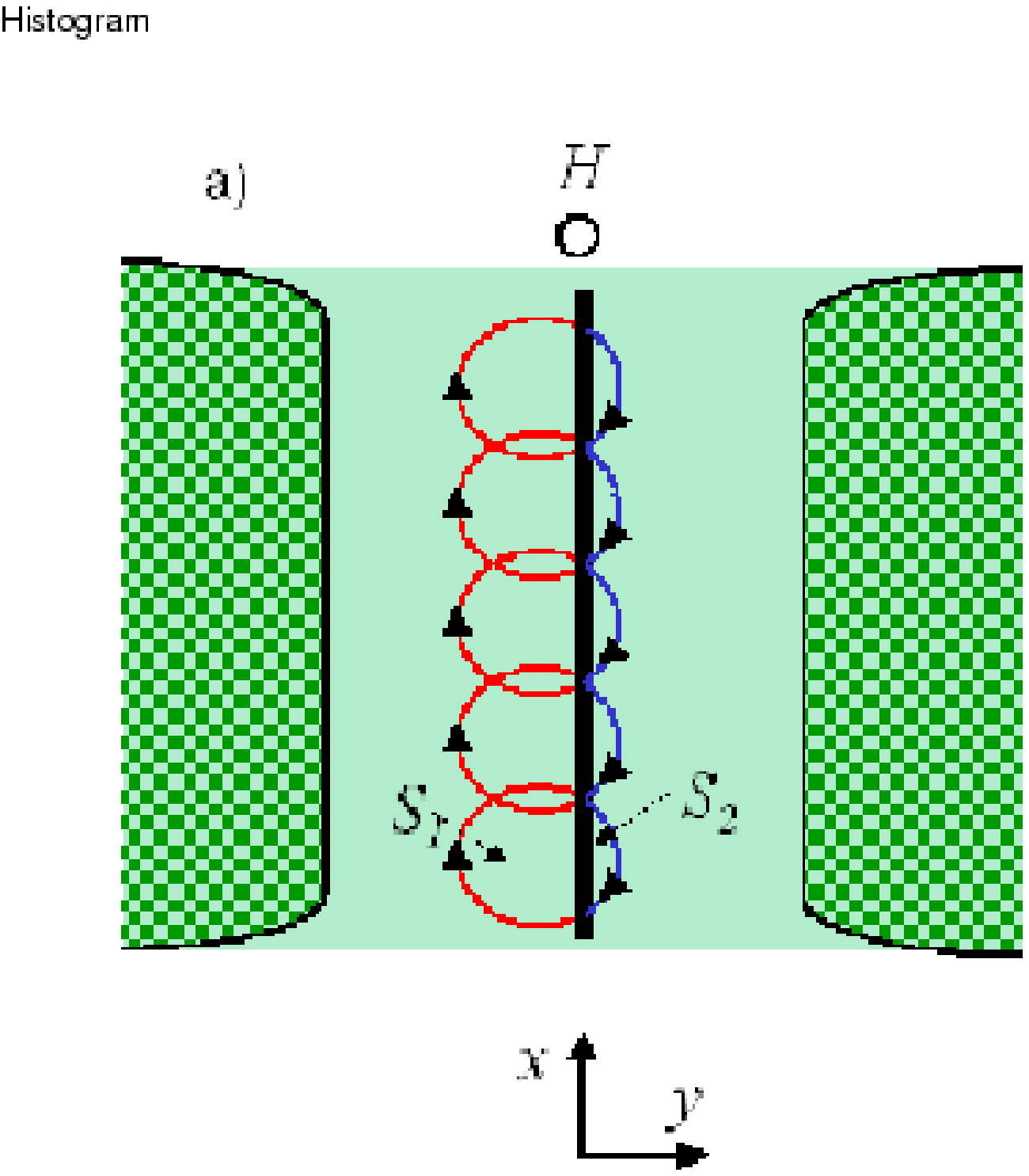} %
\includegraphics[width=3in]{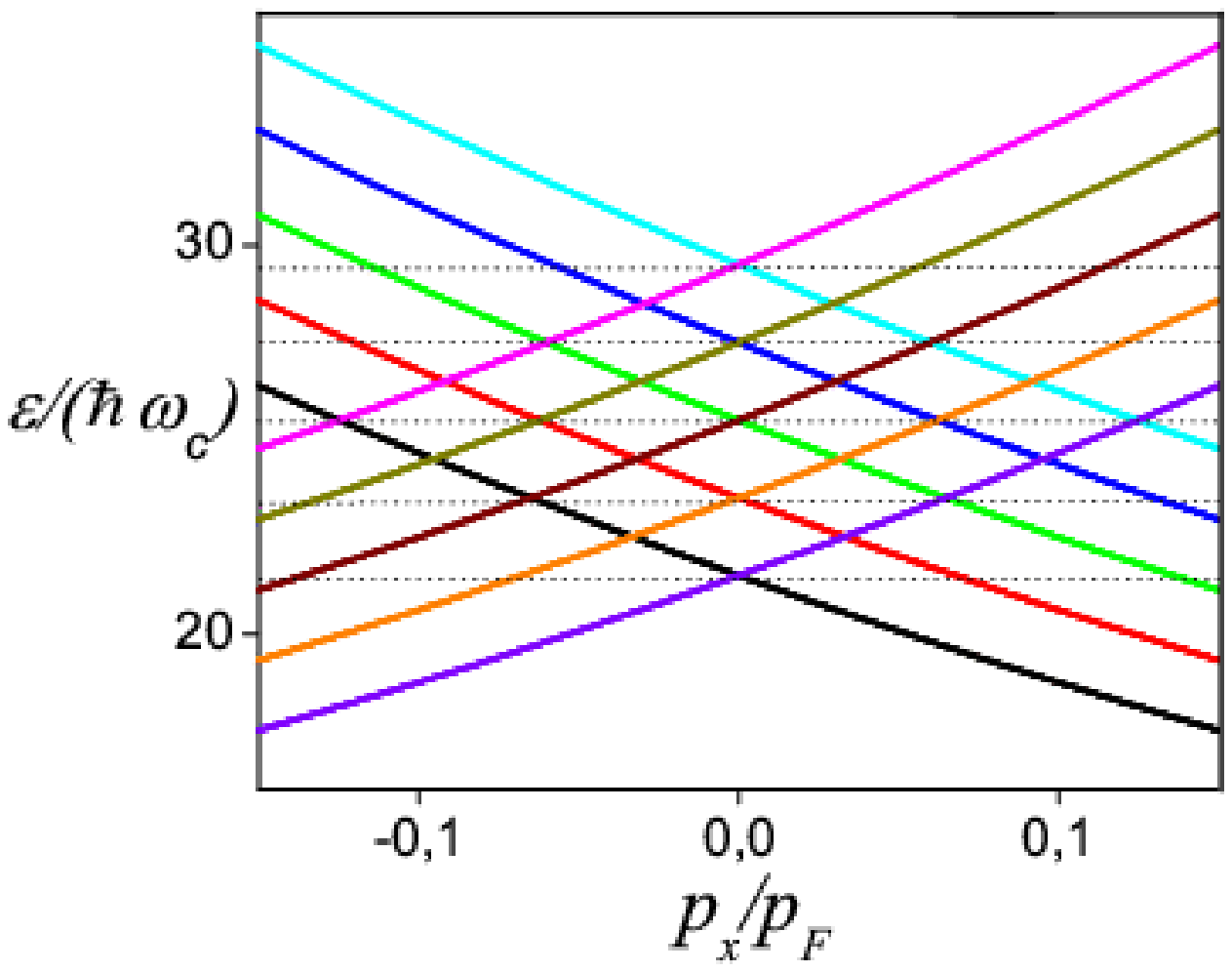}
\caption{ a) Schematic of a two-dimensional electron gas in an external
magnetic field $H$ applied perpendicular to the electron gas confinement
plane. The longitudinal barrier together with two edge states are shown. $%
S_1 $ and $S_2$ are the areas of left and right edge states.
\newline b)A part of the spectrum in the case of a zero barrier
transmission, $r=1$.
Dashed lines are degenerate Landau levels, i.e. the spectrum of the system at $%
r=0$. We use the quasiclassical parameter $\protect\alpha~=~25$. }
\label{schematic}
\end{figure}

The case of a not complete barrier transparency ($0<r<1$) is of
the most interest. In this case, the \textit{quantum interference}
between the edge states lifts the degeneracy in the crossing
points, and narrow ``energy bands'' and ``energy gaps'' appear in
the electron spectrum $\epsilon _{n}(p_{x})$. For $r\sim 1$ the
characteristic widths of the energy bands $\Delta \epsilon $ and
the energy gaps $\Delta _{g}$ are of the order $\hbar \omega
_{c}\ll \epsilon _{F}$. Notice that the electron states in the
bands are delocalized, and thus, electrons move along the barrier
with the velocity $v=d\epsilon _{n}(p_{x})/dp_{x}\sim \Delta
\epsilon /\delta p_{x}\sim v_{F}$ ($n$ is the band number).

Next, we analyze quantitatively the spectrum of the system. For
this purpose we solve the Schr\"{o}dinger equation for a
two-dimensional electron gas (in the plane ($x,y$)) in the
presence of a magnetic field $H$ and of the barrier described by a
potential $V(y)$:
\begin{equation}
\left\{ \frac{1}{2m}\left( -i\hbar \frac{\partial }{\partial x}+\frac{eHy}{c}%
\right) ^{2}-\frac{\hbar ^{2}}{2m}\frac{\partial ^{2}}{\partial y^{2}}%
+V(y)-\epsilon \right\} \Psi =0  \label{Shrod}
\end{equation}%
We assume that the characteristic width of the barrier $l_{b}$ is much
smaller than the cyclotron radius, $l_{b}\ll R_{c}$.

In the absence of the magnetic field the scattering of electrons by the
potential barrier is described by a $2\times 2$ scattering matrix

\begin{equation}
\widehat{\rho }=i e^{i\varphi }\left(
\begin{tabular}{ll}
$\ |r|$ $e^{-i\chi }$ & \hspace{0.2cm} $t$ \\
$\hspace{0.2cm}-t^*$ & $|\ r|$ $e^{i\chi }$%
\end{tabular}%
\ \right) ;\ \ \ |r|^{2}+|t|^{2}=1,\ \   \label{scatt}
\end{equation}%
where $r$ and $t$ are the probability amplitudes for an incident
electron to be reflected back and to be transmitted through the
barrier; $\varphi $ and $\chi $ are the scattering phases.

Solving Eq.(\ref{Shrod}) in the quasi-classical approximation at
distances $|y|\gg l_{b},$ where the potential barrier is
negligibly small, and matching
the wave functions of the electron on the left ($y<0$) and the right ($y>0$%
) sides of the barrier
with the help of the scattering matrix, Eq.(\ref{scatt}), we come
to the following dispersion equation \cite{comment1}:
\begin{equation}
D \equiv \cos (\frac{\pi \Phi _{+}(\varepsilon)}{2\phi _{0}}+\varphi )-r\cos (\frac{%
\pi \Phi _{-}(\varepsilon ,p_{x})}{2\phi _{0}}+\chi )=0~,
\label{spectrum}
\end{equation}%
where $\phi _{0}=hc/2e$ is the flux quantum, $\Phi _{\pm }=HS_{\pm
}$, $S_{\pm }=S_{1}\pm S_{2}$, and $S_{1,2}$ are two areas bounded
by the electron orbits (see Fig. 1a). Although  Eq.
(\ref{spectrum}) is valid for an arbitrary dispersion relation of
electrons, it becomes an extremely transparent for the parabolic
spectrum of electrons: the complete orbit is a circle with the
radius $R=c\sqrt{2m\varepsilon}/(eH)$ and the centrum shifted on
the distance $cp_x/eH$ along the $y$-axis (see Fig. 1a).

The spectrum $\varepsilon _{n}(p_{x})$ obtained from Eq.
(\ref{spectrum})
depends on both the electron momentum $p_{x}$ and a discrete quantum number $%
n$ (the band number). It displays  gaps and bands with an almost
periodic dependence in a wide region of $p_{x}$ (see Fig. 2a). In
this sense it resembles the energy spectrum of electrons in
semiconducting superlattices. However, in our case the spectrum
can be tuned by an external magnetic field and/or the reflection
coefficient $r$ controlled by the gate voltage. Moreover, the
energy levels $\varepsilon_n$ for a fixed value of $p_x$ are
distributed in a pseudo-random way. The typical distribution of
energy levels is presented in Fig. 2b.

Using Eq. (\ref{spectrum}) we calculate the electron density of
states (DOS) $\nu \left( \varepsilon \right) $ and the conductance
$\sigma \left( \varepsilon_F \right) $. Of course, this gives only
the contribution of the edge states bound to the longitudinal
barrier. If the energy $\varepsilon $ is located between the
Landau levels in the bulk, another well known contribution comes
from the edge states on the boundaries of the sample. However,
this contribution is a smooth function of the energy and is not
interesting for us.

The electron DOS $\nu (\varepsilon)=(2 \pi \hbar )^{-1}\int d p_x
\sum_n \delta (\varepsilon - \varepsilon _n(p_x))$ can be written
in the form  $\nu (\varepsilon) =(2 \pi \hbar )^{-1}\int d p_x
|\partial D/\partial \varepsilon| \delta (D)$ (see
Eq.(\ref{spectrum})); expanding it into Fourier series in
$\Phi_-(\varepsilon, p_x)$ and dropping terms fast oscillating in
$p_x$  one finds that the main contribution to DOS comes from the
zero harmonic \cite{Kaganov}. Calculating it one gets the electron
DOS  in the following form:
\begin{eqnarray}
\nu (\varepsilon)= \frac{\sqrt{2m\varepsilon}T_+}{\pi^3\hbar^2}
\frac{\left|\sin{\Phi(\varepsilon)}\right|}{\sqrt{r^{2}-\cos{\Phi(\varepsilon)}}}
\theta \left(r^{2}-\cos^2{\Phi(\varepsilon)} \right)
 \label{density}
\end{eqnarray}
where  $\Phi(\varepsilon) = 2\pi \varepsilon/(\hbar
\omega_c)+\varphi $ and  $T_{+}$ is the period of  electron motion
along the closed orbit for a given $\varepsilon$ and $\theta (x)$
is the step function. One can see from Eq. (\ref{density}) that
there are gaps in the DOS which can be found from the condition
$\cos ^{2}{(2\pi \varepsilon/\hbar \omega_c +\varphi )}>r^{2}$.

 Such a
dramatic transformation of the electron spectrum has to lead to
various novel effects in transport properties of 2DEG. As an
example, we analyze next the ballistic transport along the $x$
direction. With the standard Landauer approach based on the
relationship between the conductance and the transmission
probability in propagating channels \cite{MestranspGen2}, and
performing calculations identical to those for Eq.(\ref{density}),
we obtain the dependence of the linear conductance on the value of
the Fermi energy level $\varepsilon _{F}$
\begin{equation}
G= \frac{e^{2}}{2 h }\frac{\varepsilon_F}{\hbar
\omega_c}\sum_{n}\left(\tanh{\frac{\varepsilon_n^{(t)} -
\varepsilon_F}{2 T}}-\tanh{\frac{\varepsilon_n^{(b)}  -
\varepsilon_F}{2 T}} \right)\label{conductance}
\end{equation}
where $\varepsilon_n^{(t)}=(n\pi +\pi -\arccos{r}) \hbar
\omega_c/2$ and $\varepsilon_n^{(b)}=(n\pi +\arccos{r}) \hbar
\omega_c/2$ are the top and the bottom of the $n$ energy band,
respectively.

\begin{figure}[tbp]
\includegraphics[width=3in]{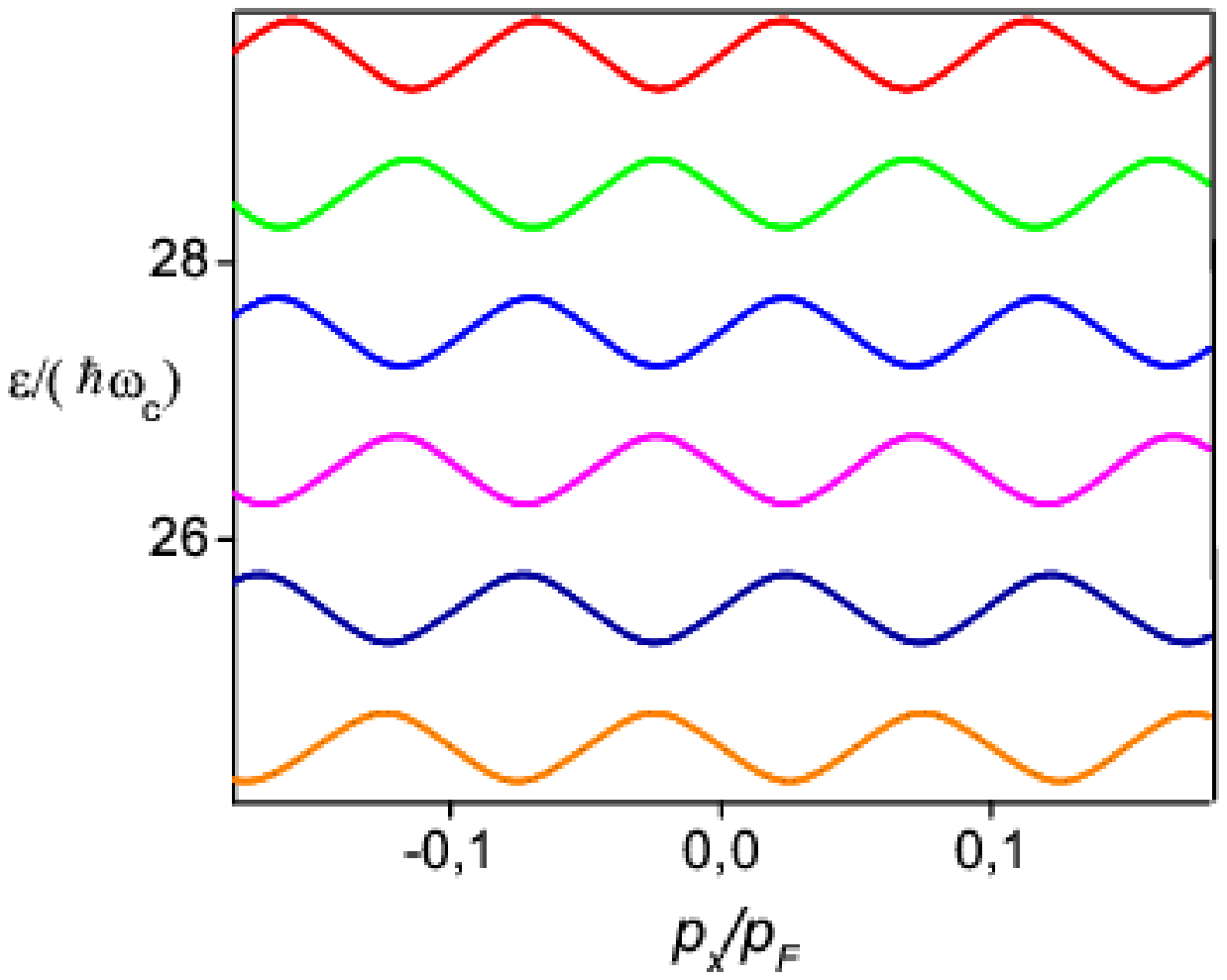} %
\includegraphics[width=2.5in]{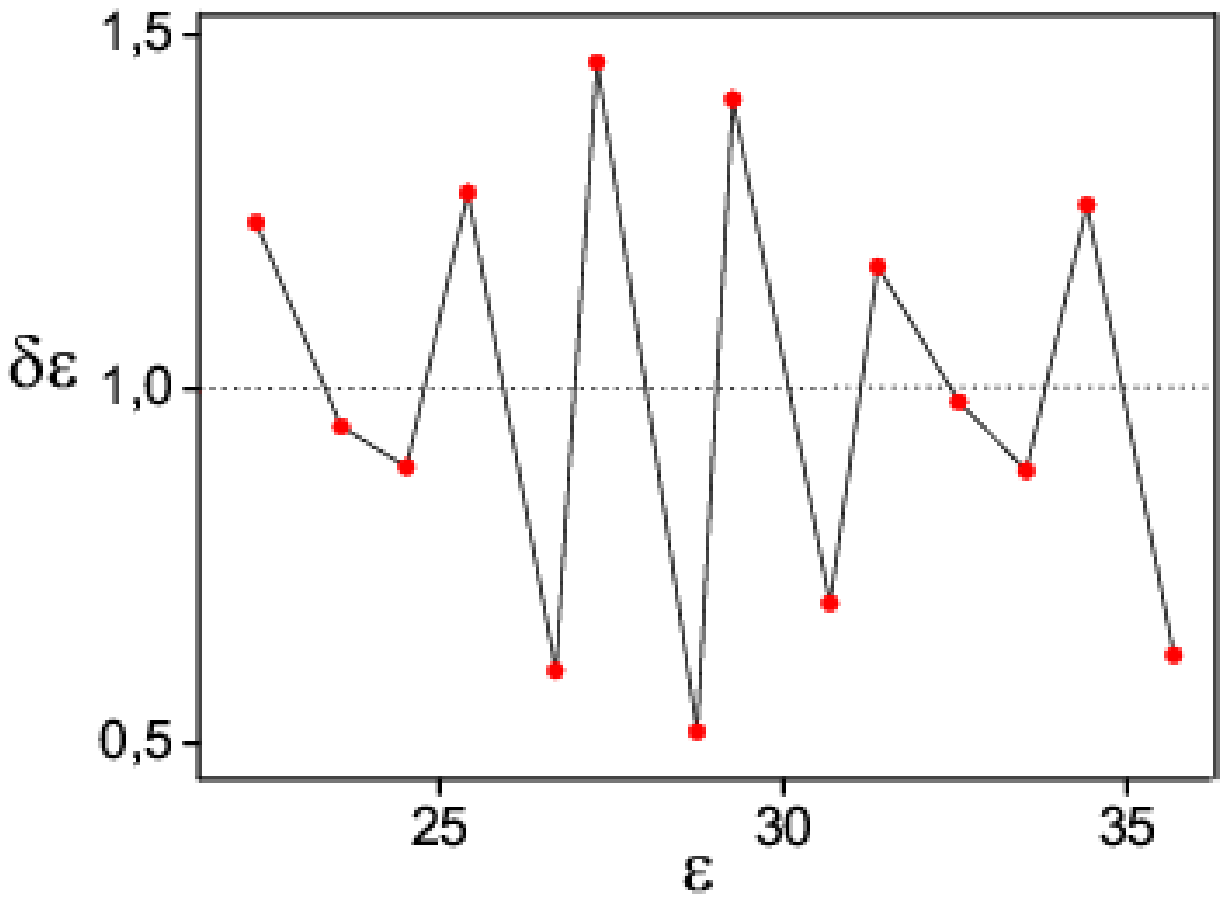}
\caption{ a) A part of the spectrum in the case of an intermediate
barrier transmission, $r=0.7$. The phases $\protect\varphi $ and
$\protect\chi $ are
obtained for the model of the $\protect\delta $-function barrier, $V(y)=V\protect%
\delta (y)$. \newline b) The dependence of the energy level
difference $\delta \protect\varepsilon = \varepsilon_{n+1} -
\varepsilon_{n}$ on the energy $\protect\varepsilon_n $ (energy
levels distribution). We use the
quasiclassical parameter $\protect\alpha ~=~25$, and a particular value of $%
p_{x}/p_F~=~0.3$. The points are connected by a thin line just for
clarity. } \label{Spectrum}
\end{figure}

The conductance $G $, Eq. (\ref{conductance}), becomes very
sensitive to the Fermi energy $\varepsilon _{F}$. Experimentally,
this energy can be tuned by applying a gate voltage.
The typical dependence of $G $ on the gate voltage displaying
\emph{giant oscillations of the conductance} is shown in Fig. 3.
These oscillations of the conductance reflect the presence of the
bands and gaps in the spectrum of the edge states, thus proving
the quantum interference of the edge states. The oscillations are
smeared by temperature (compare two curves in Fig. 3).

\begin{figure}[tbp]
\includegraphics[width=2.5in]{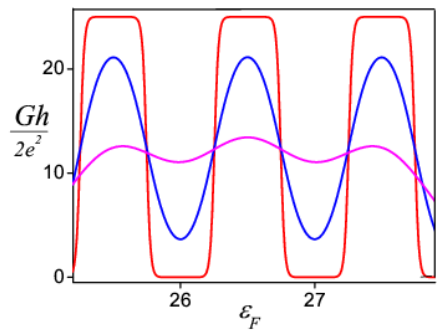}
\caption{ The giant oscillations of the conductance as a function
of the Fermi energy (the gate voltage). The influence of
temperature is shown: $k_B T_1=0.03 (\hbar \omega_c)$,  $k_B
T_2=0.2 (\hbar \omega_c)$ and $k_B T_3=0.5 (\hbar \omega_c)$. We
use the quasiclassical parameter $\protect\alpha~=~25$ and the
reflection amplitude $r=0.7$. } \label{Conductance}
\end{figure}

The oscillations found here resemble those predicted \cite{Schon} and
observed \cite{Ustinov} in the conductance of a junction between a
superconductor and a two-dimensional electron gas. However, in that case the
quantum interference between hole and electron edge states occured due to
Andreev reflection on the boundary.

As the main features of the electron motion under consideration
are due to the quantum interference, thermodynamics and transport
properties of the system are extremely sensitive to the influence
of weak external fields \cite{Kaganov}. Thus, in the ballistic
regime an increase of the transport voltage should lead to  "giant
steps" in the current-voltage characteristics (CVC). The width of
the voltage steps  is determined by the width of the gaps in the
electron spectrum, i.e. $\Delta V~\simeq~\Delta_g/e$.

In the diffusive regime, using an analogy with the electron
transport in metals under magnetic break-down \cite{Sl}, a twinned
plate \cite{Kad} and semiconducting superlattices
\cite{Blochoscill} one can expect \textit{coherent Bloch
oscillations}, and hence, an $N-$ type non-linear CVC under
relatively weak electric fields. Indeed, in the
presence of an electric field $\mathcal{E}$ satisfying an inequality $\mathcal{E%
}>\hbar \omega _{c}/(el_{0})$, where $l_{0}$ is the electron mean free path,
the periodicity of $E_{n}(p_{x})$ as a function of $p_{x}$ results in
localization of electrons along the $x$-direction. The localization length
can be estimated as $L_{loc}=v_{F}\tau _{loc}\sim v_{F}\delta p_{x}/(e%
\mathcal{E})\sim \hbar v_{F}/(e\mathcal{E}R_{c})$. The
conductivity $G $ is obtained as $\sigma =n_e e^{2}u$ where $n_e$
is the density of the charge carriers and the mobility $u$ is
determined by the Einstein relation $u=D/\varepsilon _{F}$ ($D$ is
the diffusion constant). In the case under consideration the
particle
moves over the distance $L_{loc}$ during the mean free time $\tau
_{0}$, and hence $L_{loc}^{2}\sim D\tau_{0}$. Therefore, the
current carried by the localized electrons is
\begin{equation}
j=B\sigma _{0}\left( \frac{\hbar }{\tau _{0}e\mathcal{E}R_{c}}\right) ^{2}%
\mathcal{E}=B\frac{\sigma _{0}}{\mathcal{E}}\left( \frac{\hbar
}{e\tau _{0}R_{c}}\right) ^{2}  \label{CVC}
\end{equation}%
where $B$ is a constant of order unity and $\sigma_{0}$ is the
Drude conductivity at $H=0$.

Finally, we note that in the conventional situation (in the
absence of
the longitudinal barrier), a smooth (on the scale of the Fermi wave-length $%
\lambda _{F}\sim \hbar /p_{F}$) random scattering potential
$U(x,y)$ does not change quasi-classical transport properties of
2DEG and is not usually seen in experiments. In contrast, in the
presence of the longitudinal barrier the same smooth potential can
qualitatively change the electron motion along the barrier.
Arguments analogous to those presented for the Bloch oscillations
lead to the conclusion that the electrons are {\it localized} by
this potential (or by an inhomogeneity of the magnetic field
\cite{trap}) at a localization length $L_{trap}\sim (\Delta
\varepsilon/U_{0})L_{rand}$  ($U_{0}$ and $L_{rand}$ are the
characteristic value and the correlation radius of the random
potential $U$, accordingly) as soon as $\Delta \varepsilon \ll
U_0$. Therefore, the giant oscillations (Fig. 3) can  be observed
if $U_0 < \Delta \varepsilon \sim \hbar \omega_c$. Actually, this
is the condition for the observation of Schubnikov-de Haas
oscillations and the integer quantum Hall effect.

In conclusion, we demonstrated that a two-dimensional electron gas
in the presence of a magnetic field and a longitudinal barrier is
a very interesting object. A crucial property of such a system is
the quantum interference of electron edge states propagating along
the barrier that gives rise to narrow energy bands and gaps in the
electron spectrum. The spectrum is characterized by a
quasi-periodic dependence of $\varepsilon_{n}(p_{x})$ (with the
period $\sim \hbar /R_{c}\ll p_F$) in a wide region of $p_{x}$.
The  widths of the bands and the gaps  can be tuned by the
magnetic field and the gate voltage. Many interesting novel
effects in the electron transport such as  giant oscillations of
the ballistic conductance of 2DEG as a function of the gate
voltage, non-linear CVC, etc. are possible in such a system.

The authors thank the financial support of SFB 491 and A. M.
Kadigrobov gratefully acknowledges the hospitality of the TP III
Institute, Ruhr-Universit\"{a}t, Bochum.


\begin{thebibliography}{99}

\bibitem{MestranspGen1} Th. Heinzel,
\emph{Mesoscopic Electronics in Solid State Nanostructures},
Wiley-VCH (2003).

\bibitem{MestranspGen2} Th. Dittrich, G-L. Ingold, G. Sch\"on, P. H\"anggi,
B. Kramer, and W. Zwerger, \emph{Quantum Transport and Dissipation},
Wiley-VCH (1998).

\bibitem{Beenakker} C. W. J. Beenakker and H. van Houten, Solid State
Physics, \textbf{44} 1 (1991)


\bibitem{Marcus} J. A. Folk, S. R. Patel, S. F. Godijn, A. G. Huibers, S. M. Cronenwett, C. M. Marcus,
K. Campman, and  A. C. Gossard, Phys. Rev. Lett. \textbf{76,} 1699
(1996); A. M. Chang, H. U. Baranger, L. N. Pfeiffer, K. W. West,
and T. Y. Chang, Phys. Rev. Lett. \textbf{76}, 1695 (1996).

\bibitem{Klarke} M. H. Devoret , J. M. Martinis, and J. Clarke, Phys. Rev.
Lett. \textbf{55} 1908 (1985); J. M. Martinis, M. H. Devoret, and J. Clarke,
Phys. Rev. B \textbf{35}, 4682 (1987).


\bibitem{Technology} C.C. Eugster, J.A. del Alamo, M.J. Rooks, and M.R.
Melloch Appl. Phys. Lett., \textbf{64} 3157 (1994); O. M. Auslaender, A.
Yacoby, R. de Picciotto, K. W. Baldwin, L. N. Pfeiffer, and K. W. West,
Science, \textbf{295} 825 (2002).

\bibitem{comment1} In the general case, if we do not use the quasi-classical
condition, matching two independent solutions of the Schr\"odinger
equation at the barrier leads to the dispersion equation in the
form:

$V\sqrt{\frac{\hbar R_c}{p_F}}D_\varepsilon (p_x
\sqrt{\frac{R_c}{\hbar p_F}})D_\varepsilon (-p_x
\sqrt{\frac{R_c}{\hbar p_F}})=1$, where $D_\varepsilon (x)$ is the
parabolic cylinder function, and $V$ is determined by the
properties of the barrier.

\bibitem{Schon} H. Hoppe, U. Z\"ulicke, and G. Sch\"on Phys. Rev. Lett.
\textbf{84}, 1804 (2000)

\bibitem{Ustinov} D. Uhlisch, S. G. Lachenmann, Th. Sch\"{a}pers, A. I.
Braginski, H. L\"{u}th, J. Appenzeller, A. A. Golubov, and A. V. Ustinov
Phys. Rev. B \textbf{61}, 12463 (2000).

\bibitem{Kaganov} An analogous situation takes place in bulk
metals under magnetic break-down, see review paper M.I. Kaganov
and A.A. Slutskin, Physics Reports {\bf 98}, 189 (1983).

\bibitem{Sl} A.A. Slutskin, A.M. Kadigrobov
JETP Letters \textbf{28}, 201 (1978).

\bibitem{Kad} A. M. Kadigrobov and I. V. Koshkin
Sov. J. Low Temp. Phys. \textbf{12}, 249 (1986).

\bibitem{Blochoscill} C. Waschke, H. G. Roskos, R. Schwedler, K. Leo, H. Kurz, and K. K\"ohler
Phys. Rev. Lett. \textbf{70}, 3319 (1993).





\bibitem{trap} A.A. Slutskin, A.M. Kadigrobov, JETP Lett.
\textbf{8}, 17 (1968).

\end{thebibliography}
\end{document}